\documentclass[11pt,a4paper]{article}
\usepackage{amsmath,amssymb}
\usepackage{epsfig,graphicx}
\usepackage{cite}

\begin{document}

\title{
\begin{flushright}
{\normalsize Yaroslavl State University\\
             Preprint YARU-HE-07/04} \\[20mm]
\end{flushright}
{\bf PLASMA INDUCED NEUTRINO SPIN FLIP \\ 
VIA THE NEUTRINO MAGNETIC MOMENT}
}
\author{A.~V.~Kuznetsov$^a$\footnote{{\bf e-mail}: avkuzn@uniyar.ac.ru},
N.~V.~Mikheev$^{a}$\footnote{{\bf e-mail}: mikheev@uniyar.ac.ru}
\\
$^a$ \small{\em Yaroslavl State (P.G.~Demidov) University} \\
\small{\em Sovietskaya 14, 150000 Yaroslavl, Russia}
}
\date{}

\maketitle

\begin{abstract}
The neutrino spin flip radiative conversion 
processes $\nu_L \to \nu_R + \gamma^\ast$ and 
$\nu_L + \gamma^\ast \to \nu_R$ in medium are considered. 
It is shown in part that an analysis of the so-called spin light of 
neutrino without a complete taking account of both the neutrino and 
the photon dispersion in medium is physically inconsistent.
\end{abstract}

\vfill

\begin{center}
 {\it Based on the talk presented } \\
 {\it at the 13th Lomonosov Conference on Elementary Particle Physics, } \\
 {\it Moscow State University, Moscow, Russia, August 23-29, 2007}
\end{center}


\newpage

\def\D{\mathrm{d}} 
\def\E{\mathrm{e}}
\def\I{\mathrm{i}}

\section{Introduction}
\label{sec:Introduction}
The most important event in neutrino physics of the last decades was 
the solving of the Solar neutrino problem. 
The Sun appeared in this case as a natural laboratory for investigations 
of neutrino properties. 
There exists a number of natural laboratories, the supernova explosions, where 
gigantic neutrino fluxes define in fact the process energetics. 
It means that microscopic neutrino characteristics, such as the neutrino 
magnetic moment, etc., would have 
a critical impact on macroscopic properties of these astrophysical events.

One of the processes caused by the photon interaction with the neutrino 
magnetic moment, which could play an important role in astrophysics, is
the radiative neutrino spin flip transition $\nu_L \to \nu_R \gamma$. 
The process can be kinematically allowed in medium due to its influence 
on the photon dispersion, $\omega = |{\bf k}|/n$ (here $n \ne 1$ is the 
refractive index), when the medium provides the condition $n > 1$. 
In this case the effective photon mass squared is negative, 
$m_\gamma^2 \equiv q^2 < 0$. This corresponds to the well-known 
effect of the neutrino Cherenkov radiation~\cite{Grimus:1993}.  
 
There exists also such a well-known subtle effect as the additional energy $W$ 
acquired by a left-handed neutrino in plasma. 
This additional energy was considered in the 
series of papers by Studenikin et al.~\cite{0611100} as a new 
kinematical possibility to allow the radiative neutrino transition 
$\nu_L \to \nu_R \gamma$. 
The effect was called the ``spin light of neutrino''. 
For some reason, the photon dispersion in medium providing in part the 
photon effective mass, was ignored in these papers. 
However, it is evident that a kinematical analysis based on the additional 
neutrino energy in plasma (caused by the weak forces) 
when the plasma influence on the photon dispersion 
(caused by electromagnetic forces) is ignored, cannot be considered as 
a physical approach. 
In this paper, we perform a consistent analysis of the radiative neutrino 
spin flip transition in medium, when its influence both on the photon and 
neutrino dispersion is taken into account. 
\section{Cherenkov process $\nu_L \to \nu_R \gamma$ and its crossing 
$\nu_L \gamma \to \nu_R$}
\label{sec:Cherenkov}
Let us start from the Cherenkov process of the photon creation 
by neutrino, $\nu_L \to \nu_R \gamma$, which should be appended 
by the crossed process of the photon absorption 
$\nu_L \gamma \to \nu_R$. 
At this stage we neglect the additional left-handed neutrino energy $W$, 
which will be inserted below. 
For the $\nu_L \to \nu_R$ conversion width one obtains by the standard way:
\begin{eqnarray}
\Gamma^{\rm tot}_{\nu_L \to \nu_R} &=& 
\Gamma_{\nu_L \to \nu_R \gamma} +
\Gamma_{\nu_L \gamma \to \nu_R}
= \frac{\mu_\nu^2}{16 \, \pi^2 E} \int j_\alpha \, j^\ast_\beta \,
\sum\limits_{\lambda = t, \ell}  
\varepsilon_{(\lambda)}^{\ast \alpha} \, \varepsilon_{(\lambda)}^\beta 
\, Z_\gamma^{(\lambda)}
\, \frac{{\D}^3 {\bf p'}}{E'}
\nonumber\\[2mm]
&\times& \left\{
\frac{\delta (E - E' - \omega)}{2 \, \omega} 
\left[ 1 + f_\gamma (\omega) \right] +
\frac{\delta (E - E' + \omega)}{2 \, \omega} \, 
f_\gamma (\omega) 
\right\} ,
\label{eq:Gamma1}
\end{eqnarray}
where 
$\varepsilon_{(\lambda)}^{\alpha}$ is the photon 
polarization vector, $j^\alpha$ is the Fourier transform of the neutrino 
magnetic moment current, 
$p^\alpha = (E, {\bf p})$, $p '^\alpha = (E ', {\bf p} ')$ 
and $q^\alpha = (\omega, {\bf k})$ are the four-momenta of the initial and 
final neutrinos and photon, respectively, 
$\lambda = t, \ell$ mean transversal and longitudinal photon polarizations, 
$f_\gamma (\omega) = \left(\E^{\omega/T} - 1\right)^{-1}$ is the 
Bose--Einstein photon distribution function,
and 
$Z_\gamma^{(\lambda)} = (1 - \partial \Pi_{(\lambda)}/\partial \omega^2)^{-1}$ is the 
photon wave-function renormalization. 
The functions $\Pi_{(\lambda)}$, defining the photon dispersion law: 
\begin{eqnarray}
\omega^2 - {\bf k}^2 - \Pi_{(\lambda)} (\omega, {\bf k}) = 0 \,,
\label{eq:disp}
\end{eqnarray}
are the eigenvalues of the photon polarization tensor: 
$\Pi_{\alpha \beta} \, \varepsilon_{(\lambda)}^\beta 
= \Pi_{(\lambda)} \, \varepsilon_{(\lambda)\alpha}$. 

The width $\Gamma^{\rm tot}_{\nu_L \to \nu_R}$ can be rewritten 
to another form. Let us introduce the energy transferred from neutrino:
$ E - E' = q_0$, which is expressed via the photon energy 
$\omega (k)$ as $q_0 = \pm \omega (k)$. Then 
$\delta$-functions in Eq.~(\ref{eq:Gamma1}) can be rewritten: 
\begin{eqnarray}
\frac{\delta \left(q_0 \mp \omega (k) \right)}{2 \,\omega (k)} =  
\delta \left(q_0^2 - \omega^2 (k) \right) \, \theta (\pm q_0). 
\label{eq:delta1}
\end{eqnarray}
Transforming the $\delta$-function to have the 
dispersion law in the argument: 
\begin{eqnarray}
\delta \left(q_0^2 - \omega^2 (k) \right) = 
\left[ Z_\gamma^{(\lambda)} \right]^{-1} \, 
\delta \left(q^2 - \Pi_{(\lambda)} (q) \right) ,
\label{eq:delta2}
\end{eqnarray}
one can see that the renormalization factor 
$Z_\gamma^{(\lambda)}$ is cancelled in the conversion width~(\ref{eq:Gamma1}).
Integration in Eq.~(\ref{eq:Gamma1}) with 
the $\delta$-function~(\ref{eq:delta2})
can be easily performed when the function 
$\Pi_{(\lambda)} (q)$ is real.
However, it has, in general, an imaginary part.
It means, that the photon is unstable in plasma. 
\section{Generalization to the case of unstable photon}
\label{sec:unstable}
In the case, when the eigenfunction $\Pi_{(\lambda)} (q)$ has an imaginary 
part, one should use instead of the $\delta$-function its natural 
generalization of the Breit--Wigner type, 
with e.g. the retarded functions $\Pi_{(\lambda)} (q)$ :
\begin{eqnarray}
\delta \left(q^2 - \Pi_{(\lambda)} (q) \right) \; \Rightarrow \;
\frac{1}{\pi} \; \frac{- {\mathrm{Im}} \,\Pi_{(\lambda)} \; {\mathrm{sign}} (q_0) 
\, \epsilon_{\lambda}}
{\left(q^2 - \mathrm{Re} \,\Pi_{(\lambda)} \right)^2 + 
\left(\mathrm{Im} \,\Pi_{(\lambda)} \right)^2} \,,
\label{eq:Breit-Wigner}
\end{eqnarray}
where $\epsilon_{\lambda} = + 1$ 
for $\lambda = t$
and $\epsilon_{\lambda} = - 1$ 
for $\lambda = \ell$.

After some transformations, taking into account the additional energy $W$ 
acquired by a left-handed neutrino in plasma, and changing the integration 
variables from the final neutrino 3-momentum to the 
photon energy and momentum, ${\D}^3 {\bf p'} \to {\D} q_0 \, {\D} k 
\; ( k \equiv |{\bf k}|)$, 
one obtains: 

\begin{eqnarray}
&&\Gamma^{\rm tot}_{\nu_L \to \nu_R}
= \frac{\mu_\nu^2}{16\, \pi^2 \, E^2} \; 
\int\limits_{-\infty}^{E + W} \, {\mathrm{d}} q_0 \, 
\int\limits_{|q_0 - W|}^{2 E + W - q_0} \, \frac{{\D} k}{k} \, 
\left[ 1 + f_\gamma (q_0) \right] \, (2 E - q_0)^2 \, q^4
\nonumber\\[2mm]
&&\times 
\left\{ 
\left( 1 - \frac{k^2}{(2 E - q_0)^2} \right) 
\left[ 1 - \frac{2 q_0 W}{q^2} + \frac{8 E (E - q_0) W^2}
{q^4 \left[ (2 E - q_0)^2/k^2 - 1  \right]} \right] 
\varrho_{(t)} (q_0, k)
\right.
\nonumber\\[2mm]
&&- \left.
\left( 1 - \frac{2 q_0 W}{q^2} \right) \varrho_{(\ell)} (q_0, k) 
\right\} ,
\label{eq:Gamma2}
\end{eqnarray}
where $q^2 = q_0^2 - k^2 $, and the photon spectral density 
functions are introduced:
\begin{eqnarray}
\varrho_{(\lambda)} = 
\frac{2 \left(- \mathrm{Im} \,\Pi_{(\lambda)} \right)}
{\left(q^2 - \mathrm{Re} \,\Pi_{(\lambda)} \right)^2 + 
\left(\mathrm{Im} \,\Pi_{(\lambda)} \right)^2} \,.
\label{eq:spect_fun}
\end{eqnarray}

Our formula~(\ref{eq:Gamma2}) 
having the most general form, can be used for neutrino-photon 
processes in any optically active medium. We only need to identify 
the photon spectral density functions $\varrho_{(\lambda)}$. 
\section{Does the window for the ``spin light of neutrino'' exist?}
\label{sec:window}
\begin{figure}[htb]
\centering
\includegraphics[width=0.95\textwidth]{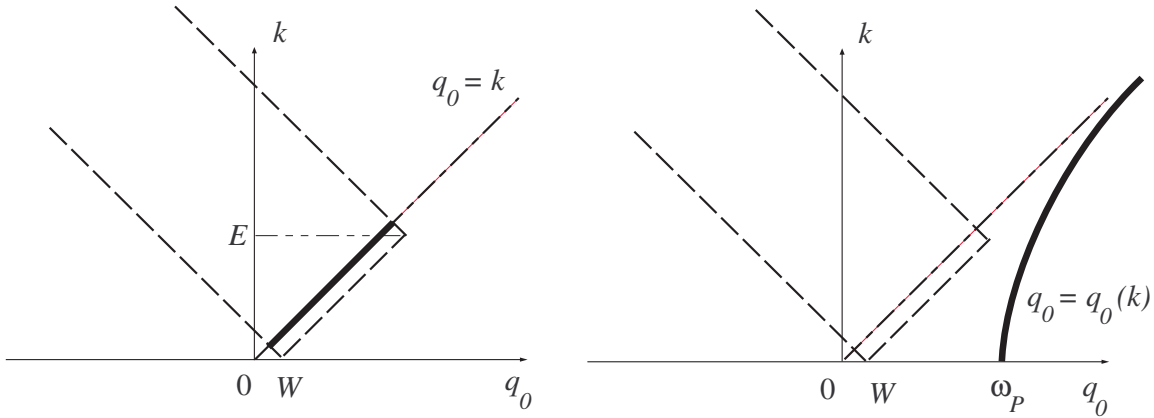}
\caption{
The region of integration for the width $\Gamma^{\rm tot}_{\nu_L \to \nu_R}$ 
with the fixed initial neutrino energy $E$ is inside the slanted rectangle 
shown by dashed line. The vacuum photon dispersion (if the medium influence 
is ignored) is shown by bold line in the left plot. 
The photon dispersion curve in plasma is shown by bold line in the right plot.
}
\label{fig:disp}
\end{figure}

To show manifestly that the case considered in the papers by 
Studenikin et al.~\cite{0611100}, with taking the additional left-handed 
neutrino energy $W$ in plasma and ignoring 
the photon dispersion, was really unphysical, let us 
consider the region of integration for the width 
$\Gamma^{\rm tot}_{\nu_L \to \nu_R}$ in Eq.~(\ref{eq:Gamma2}).  
In Fig.~\ref{fig:disp}, the photon vacuum dispersion line $q_0 = k$ 
is inside the allowed kinematical region (left plot), but 
the plasma influenced photon dispersion curve appears to be outside, 
if the neutrino energy is not large enough (right plot). 
 
For the fixed plasma parameters, the threshold neutrino energy 
$E_{\mathrm{min}}$ exists for coming of the dispersion curve into 
the allowed kinematical region. Even for the interior of a 
neutron star this threshold neutrino energy is rather large: 
$
E_{\mathrm{min}} \simeq {\omega_P^2}/{(2 \, W)} 
\simeq 10 \,{\rm TeV} \,, 
$
where $\omega_P$ is the plasmon frequency.

One could hope that the ``spin light of neutrino'' may be possible at 
ultra-high neutrino energies. However, in this case the local limit of 
the weak interaction is incomplete, and the non-local weak 
contribution into additional neutrino energy $W$ 
must be taken into account. This contribution 
always has a negative sign, and its absolute value grows with the 
neutrino energy. One could only hope that 
the window arises in the 
neutrino energies for the process to be kinematically opened, 
$E_{\mathrm{min}} < E < E_{\mathrm{max}}$. 
For example, in the solar interior there is no window 
for the process with electron neutrinos at all. 
A more detailed analysis of this subject was performed in our 
papers~\cite{Kuznetsov:2006,Kuznetsov:2007}. 

\section*{Acknowledgements}

A. K. expresses his deep gratitude to the organizers of the 
13th Lomonosov Conference on Elementary Particle Physics 
for warm hospitality.

The work was supported in part 
by the Russian Foundation for Basic Research under the Grant No. 07-02-00285-a. 



\end{document}